# Nature does not rely on long-lived electronic quantum coherence for photosynthetic energy transfer

Hong-Guang Duan,[1,2,3] Valentyn I. Prokhorenko,[1] Richard Cogdell,[4]
Khuram Ashraf,[4] Amy L. Stevens,[1] Michael Thorwart,[2,3,*] R. J. Dwayne Miller[1,3,5,*]

[1]Max Planck Institute for the Structure and Dynamics of Matter,
Luruper Chaussee 149, 22761, Hamburg, Germany.
[2]I. Institut für Theoretische Physik, Universität Hamburg,
Jungiusstraße 9, 20355 Hamburg, Germany.
[3]The Hamburg Center for Ultrafast Imaging,
Luruper Chaussee 149, 22761 Hamburg, Germany.
[4]Institute of Molecular, Cell and Systems Biology, College of Medical,
Veterinary and Life Science, University of Glasgow, Glasgow G12 8QQ, UK.
[5]The Departments of Chemistry and Physics, University of Toronto,
80 St. George Street, Toronto Canada M5S 3H6.

*To whom correspondence should be addressed;
E-mail: michael.thorwart@physik.uni-hamburg.de, dwayne.miller@mpsd.mpg.de.

**During the first steps of photosynthesis, the energy of impinging solar photons is transformed into electronic excitation energy of the light-harvesting biomolecular complexes. The subsequent energy transfer to the reaction center is understood in terms of exciton quasiparticles which move on a grid of biomolecular sites on typical time scales less than 100 femtoseconds (fs). Since the early days of quantum mechanics, this energy transfer is described as an incoherent Förster hopping with classical site occupation probabilities, but**

with quantum mechanically determined rate constants. This orthodox picture has been challenged by ultrafast optical spectroscopy experiments (*1–3*) with the Fenna-Matthews-Olson protein (*4*) in which interference oscillatory signals up to $1.5$ picoseconds were reported and interpreted as direct evidence of exceptionally long-lived electronic quantum coherence. Here, we show that the optical 2D photon echo spectra of this complex at ambient temperature in aqueous solution do not provide evidence of any long-lived electronic quantum coherence, but confirm the orthodox view of rapidly decaying electronic quantum coherence on a time scale of $60$ fs. Our results give no hint that electronic quantum coherence plays any biofunctional role in real photoactive biomolecular complexes. Since this natural energy transfer complex is rather small and has a structurally well defined protein with the distances between bacteriochlorophylls being comparable to other light-harvesting complexes, we anticipate that this finding is general and directly applies to even larger photoactive biomolecular complexes.

The principle laws of physics undoubtedly also govern the principle mechanisms of biology. The animate world consists of macroscopic and dynamically slow structures with a huge number of degrees of freedom such that the laws of statistical mechanics apply. On the other hand, the fundamental theory of the microscopic building blocks is quantum mechanics. The physics and chemistry of large molecular complexes immersed in solution may be considered as a bridge between the molecular world and the formation of living matter. A fascinating question since the early days of quantum theory is on the borderline between the atomistic quantum world and the classical world of biology. Clearly, the conditions under which matter displays quantum features or biological functionality are contrarious. Quantum coherent features only become apparent when systems with a few degrees of freedom with a preserved quantum me-chanical phase relation of a wave function are well shielded from environmental fluctuations that otherwise lead to rapid dephasing. This dephasing mechanism is very efficient at ambient temperatures at which biological systems operate. Also, the function of biological macromolecular systems relies on their embedding in a "wet" and highly polar solvent environment, which is again hostile to any quantum coherence. So, the common view has developed that a quantum coherent phase relation between the pigments involved in any biologically relevant dynamical process is rapidly destroyed on a sub-100 fs time scale. The transport is dominated by incoherent coupling between sites in which the energy transport is spatially directed by differences in site energies that naturally lead to energy relaxation and flow to the special pair. The nature of any quantum coherent interaction is short lived and would only involve nearest neighbours, a distinctively different view and operating physics from that of a fully coherent process as implicated by electronic coherences living as long as proposed (*1–3*).

In the recent years, ultrafast nonlinear two-dimensional (2D) optical spectroscopy has made it possible to challenge this orthodox view, since it accesses the fs time scales. A prominent energy transfer complex which is simple enough to provide clean experimental spectroscopic data is the Fenna-Matthew-Olson (FMO) protein (*4*). In 2007, the oscillatory beatings observed in the off-diagonal signals of 2D optical spectra have been reported to survive for longer than 660 fs at cryogenic temperatures at 77 K (*1*). They were interpreted as signatures of long-lived electronic quantum coherent exciton dynamics in the FMO network. The same conclusion was drawn also from an experiment at higher temperatures up to 277 K (*2*), and similar beatings have been reported for marine cryptophyte algae (*3*) as well. These experiments have triggered an enormous interest in a potential new field of "quantum biology" (*5–9*), with far reaching consequences even for the functionality of the human brain (*10*) and for technological applications (*11*). The cornerstone experiments (*1–3*) remained unconfirmed until recently, when a new effort was made to identify the excitonic energy transfer pathways (*12*) at 77 K. However, observation of long-lived electronic coherences in the 2D off-diagonal signals was not reported in Ref. 12.

Interesting conceptual studies on the possible constructive role of dephasing fluctuations on the efficiency of energy transfer in networks (*13–18*) were initiated subsequently. However, also critical questions were formulated (*19–23*) which remained largely unresolved up to present. Advanced dynamical simulations (*24, 25*) did not confirm long-lived electronic coherence. An additional strong coupling of the excitonic to the nuclear degrees of freedom was also considered (*26–33*) as a possible driving source of coherence. Such a mechanism can yield longer lived oscillations of the cross-peak amplitude (*34, 35*), yet, a strong vibronic coupling is required and the oscillation amplitudes typically remain small. Moreover, the coupling between such oscillations can only persist as long as the electronic coherence lives as long as the vibrational period, an issue that has not been discussed.

Motivated by the unsatisfactory lack of experimental confirmation of the long-lived electronic coherence at ambient temperatures, we have revisited this question in the present combined experimental and theoretical study with the important distinction that we have conducted these studies under physiologically relevant conditions. We have measured a series of photon echo 2D optical spectra of the FMO trimer, extracted from green sulfur bacteria *C. tepidum*, at room temperature (296 K or 23° C) and for different waiting times $T$. The 2D spectra for selected times between $T = 0$ and $T = 2000$ fs are shown in Fig. 1 (a). In addition, we have calculated the 2D spectra of the FMO monomer (see Supplementary Materials for details) and compared them to the experimental ones (Fig. 1 (a)). In passing, we note that the calculated absorption spectrum shown in Fig. 1 (c) coincides with that measured in Ref. 36. The experimental and calculated 2D spectra for different waiting times agree well. At initial waiting time $T = 0$ the 2D spectrum is stretched along the diagonal, which is a manifestation of a significant inhomogeneous broadening. With increasing waiting time, the inhomogeneous broadening is rapidly reduced and becomes undetectable beyond $T \sim 1000$ fs. A central feature of 2D electronic spectroscopy is that the antidiagonal width uniquely reveals the electronic dephasing time scale. This also holds when a strong vibronic coupling is present (*34*). In Fig. 1 (b), we show the antidiagonal cut of the 2D spectrum measured at $T = 0$ (indicated in Fig. 1 (a) by the black line) with a fitted Lorentzian profile yielding a FWHM of $\Delta_{\text{hom}} = 175$ cm$^{-1}$. This corresponds to an electronic dephasing time of $\tau_{\text{hom}} = [\pi c \Delta_{\text{hom}}]^{-1} = 60$ fs ($c$ is the speed of light), which sets a principle upper limit for the decay times of any possible oscillations originating from the beatings between the excitonic transitions. Furthermore, the magnitude of the off-diagonal peaks in the upper left region of the 2D spectra remarkably increases for growing waiting times. This renders the spectrum strongly elongated along the $\omega_\tau$ coordinate and reveals an efficient energy transfer between the FMO pigments. The progressions ranging from the central peak at $\omega_\tau = \omega_t = 12400$ cm$^{-1}$ to the region $\omega_\tau = 13500$ cm$^{-1}$, $\omega_t = 12400$ cm$^{-1}$ indicate a vibrational relaxation of the localized vibrational modes of the bacteriochlorophylls. In addition, we clearly observe a fast decay of the central peak amplitude within the first $1 - 2$ ps, induced by thermally fluctuating electric fields from the polar protein environment.

The energy transfer pathways and the associated timescales are revealed by a multidimensional global analysis (*37*) of consecutive 2D spectra at different waiting times $T$, arranged in a three-dimensional array $S(\omega_\tau, \omega_t, T)$ (for a detailed description, see Supporting Information to Ref. 38). This results in two-dimensional decay-associated spectra (2DDAS) $A_i(\omega_\tau, \omega_t)$ which are shown in Fig. 2 (a) (left column: analyzed experimental spectra, right column: analyzed theoretical data, see also Supplementary Materials for more details). We have resolved four different energy transfer timescales. The shortest decay time of $\tau_1 = 90$ fs is associated to the positive diagonal peak at $12500$ cm$^{-1}$. Furthermore, a strong negative off-diagonal peak at $\omega_\tau = 12500$ cm$^{-1}$, $\omega_t = 12000$ cm$^{-1}$ indicates the energy transfer from the excitonic states located around $12500$ cm$^{-1}$ to the lower ones at $12000$ cm$^{-1}$. The second 2DDAS, associated with a lifetime of $\tau_2 = 750$ fs, displays similar features, but with slightly broader peaks and a noticeable extension to the blue spectral side. The third ($\tau_3 = 7.0$ ps) and fourth ($\tau_4 = \infty$) components of the 2DDAS only show one diagonal peak at the central position of $12200$ cm$^{-1}$ with rather broad peaks. This indicates a thermal relaxation of the pigments inside the FMO protein. Our findings qualitatively agree with those in Ref. 12, with all our transfer rates being smaller which is due to the high temperature.

Next, we address the question of long-lived coherent oscillations in off-diagonal signals in the 2D spectra. We have analyzed the residuals obtained after removing the underlying slow kinetics from the 3D data set $S(\omega_\tau, \omega_t, T)$. Their Fourier transform provides a three-dimensional spectrum of the possible vibrations. The most intense of them with the amplitudes above the noise threshold are plotted in Fig. S9 of the Supplementary Materials. As a result of our model, all the exciton states in the FMO complex are located in the frequency region of $12123 - 12615$ cm$^{-1}$. Hence, the largest oscillation frequency which can be expected from the beatings between them is $\sim 490$ cm$^{-1}$. However, the lowest oscillation frequency which we found in the residuals lies well above ($\sim 600$ cm$^{-1}$). Hence, we can safely conclude that the origin of these oscillations is not due to interference between the excitonic states.

A cross-correlation analysis (*38*) of the residuals across the diagonal $\omega_\tau = \omega_t$ in a delay time window up to 2 ps yields a 2D correlation spectrum shown in Fig. 2 (b) where the positive (negative) values indicate (anti-)correlated residuals. We find two strong negative peaks, which proves on the basis of Ref. 39, 40 that the oscillations in this region are related to vibrational coherence. Moreover, we clearly observe two negative peaks at the frequencies $12400$ cm$^{-1}$ and $13300$ cm$^{-1}$. They can be associated to strong localized vibrational modes of the bacteriochlorophylls which follows from the vibrational progression in the absorption spectrum shown in Fig. 1 (c).

To underpin the vibrational origin of the oscillations with rather small amplitudes, we con-sider the time evolution of the two off-diagonal signals located at the pair of frequencies 12300 cm$^{-1}$ and 12600 cm$^{-1}$ marked by the red and blue squares in Fig. 1 (a). We consider the results of the theoretical modeling which on purpose only includes an overdamped vibrational mode (with a vibrational lifetime of $\tau_\Gamma = 15$ fs, see Supplementary Materials), but do not include underdamped vibrational states of the bacteriochlorophylls. Hence, any calculated oscillations occuring out to long time must originate from possible beatings of the electronic dynamics due to a coherent coupling between the excitonic states. This helps to uniquely determine the origin of the oscillations observed in the experimental spectra. The results shown in Figs. 3 (a) and (b) prove that any electronic coherence vanishes within a dephasing time window of 60 fs. It is important to emphasize that the dephasing time determined like this is consistent with the dephasing time of $\tau_{\text{hom}} = 60$ fs independently derived from the experiment (see above). It is important to realize that this cross-check constitutes the simplest and most direct test for the electronic dephasing time in 2D spectra. In fact, the only unique observable in 2D photon echo spectroscopy is the homogeneous lineshape. The use of rephasing processes in echo spectroscopies removes the inhomogeneous broadening and this can be directly inferred by the projection of the spectrum on the antidiagonal that shows the correlation between the excitation and probe fields. This check of self-consistency has not been made earlier and is in complete contradiction to the assertion made in earlier works. Moreover, our direct observation of the homogeneous line width is in agreement with independent FMO data of Ref. 12. This study finds an $\sim 100$ cm$^{-1}$ homogeneous line width estimated from the low-temperature data taken at 77 K, which corresponds to an electronic coherence time of $\sim 110$ fs, in line with our result given the difference in temperature. In fact, if any long lived electronic coherences were operating on the 1 ps timescale as claimed previously (*1*), the antidiagonal line width would have to be on the order of 10 cm$^{-1}$, and would appear as an extremely sharp ridge in the 2D inhomogeneously broadened spectrum (see Supplementary Materials). The lack of this feature conspicuously points to the misassignment of the long lived features to long lived electronic coherences where as now established in the present work is due to weak vibrational coherences. The frequencies of these oscillations, their lifetimes, and amplitudes all match those expected for molecular modes (*41, 42*) and not long-lived electronic coherences.

For a comparison with the previous experimental work (at the temperature 277 K (4° C)) of Ref. 2, we have extracted from our measured 2D spectra the time evolution of the "cross peak" located at $\omega_\tau = 12350$ cm$^{-1}$, $\omega_t = 12200$ cm$^{-1}$ (spectral position marked by a black cross in Fig. 1 (a)). As follows from Fig. 3 (c) (see also Supplementary Materials), no long-lived beatings and associated electronic coherence can be observed at this position in the measured 2D spectra. Within the available experimental signal-to-noise ratio, we can conclude that there are no oscillations with amplitudes larges than 5% of the signal. They may be interpreted as weak vibrational coherence, consistent with earlier findings (*35*) for a much smaller excitonically coupled system (a dimer) with strong vibronic coupling.

The present work, which has considered a full analysis of possible electronic state couplings, decay associated spectra, signs/amplitudes of off-diagonal features and most telling the directly determined homogeneous lineshape that the previous assignment to long lived electronic coherences is incorrect. There is no long range coherent energy transport occurring in the FMO complex and in all cases is not needed to explain the overall efficiency of energy transfer. This constitutes the main result of our work and confirms the orthodox picture of rapidly decaying electronic coherence on a time scale of 60 fs in the exciton dynamics in the FMO protein complex at ambient temperature. In turn, it disproves any contributions of quantum coherence to biological functionality under ambient conditions in natural light-harvesting units, in line with our previous study of the light-harvesting complex LHCII (*38*). Since the FMO complex is rather small and a structurally quite well defined protein with the distances between bacteriochlorophylls comparable to other natural light-harvesting systems, we anticipate that this finding in generic also applies to even larger photoactive biomolecular complexes.

**Acknowledgments**

H.-G.D. thanks Eike-Christian Schulz for his comments on the properties of the FMO protein at room temperature. We acknowledge financial support by the Max Planck Society and the Hamburg Centre for Ultrafast Imaging (CUI) within the German Excellence Initiative supported by the Deutsche Forschungsgemeinschaft. H-G.D. acknowledges generous financial support by the Joachim-Hertz-Stiftung Hamburg. Work at the University of Glasgow was supported as part of the Photosynthetic Antenna Research Center (PARC), an Energy Frontier Research Center funded by the U.S. Department of Energy, Office of Science, Basic Energy Sciences under Award #DE-SC0001035.

**Supplementary Materials**

Materials and Methods

Figs. S1 to S12

List of References

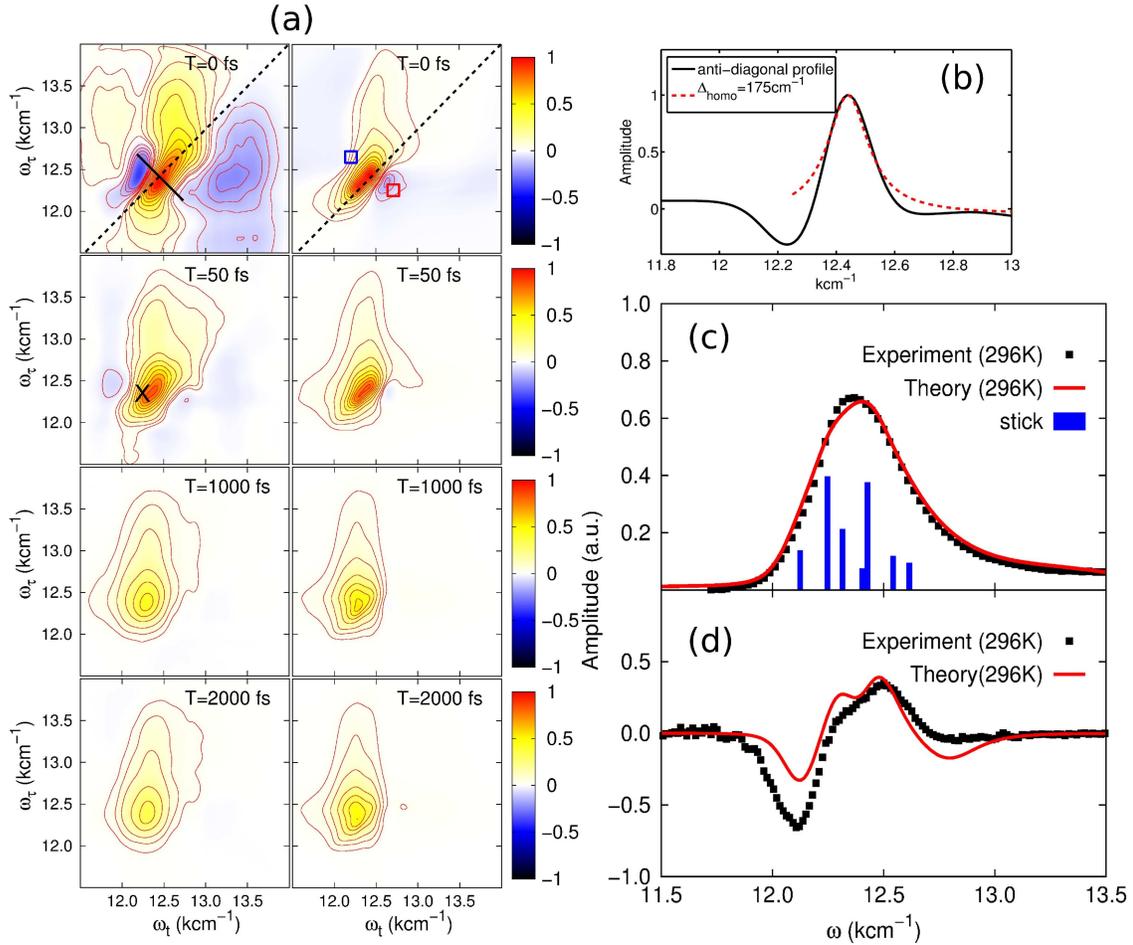

Figure 1: **Spectroscopy of the FMO complex:** (a) 2D photon echo spectra for different waiting times $T$ taken at 296 K. Shown is the real part. The left column shows the experimental results, while the right column displays the theoretically calculated spectra using the parameters obtained from the fits shown in (c) and (d). The black solid line marks the antidiagonal along which the spectral signal is shown in (b). It is used as a consistence check for the electronic dephasing time. The blue and red squares in the top right panel mark the spectral positions at which we evelute the cross-peak time evolution shown in Fig. 3 (a) and (b). The black cross in the left second panel marks the spectral position evaluated in Fig. 3 (c). (b) Spectral profile along the antidiagonal as extracted from the experimental data shown in (a). we note that the negative amplitude part in the profile shows the signal from the solvent at $T = 0$ fs, which leads to a slight underestimation of the homogeneous line width. (c) Linear absorption spectrum at 296 K. Black symbols indicate the measured data, the red solid line marks the theoretically calculated result. The blue bars show the calculated stick spectrum. (d) Circular dichroism spectrum at 296 K. Black symbols mark the measured data. The red solid lines shows the theoretical result calculated with the same parameters as in (c).

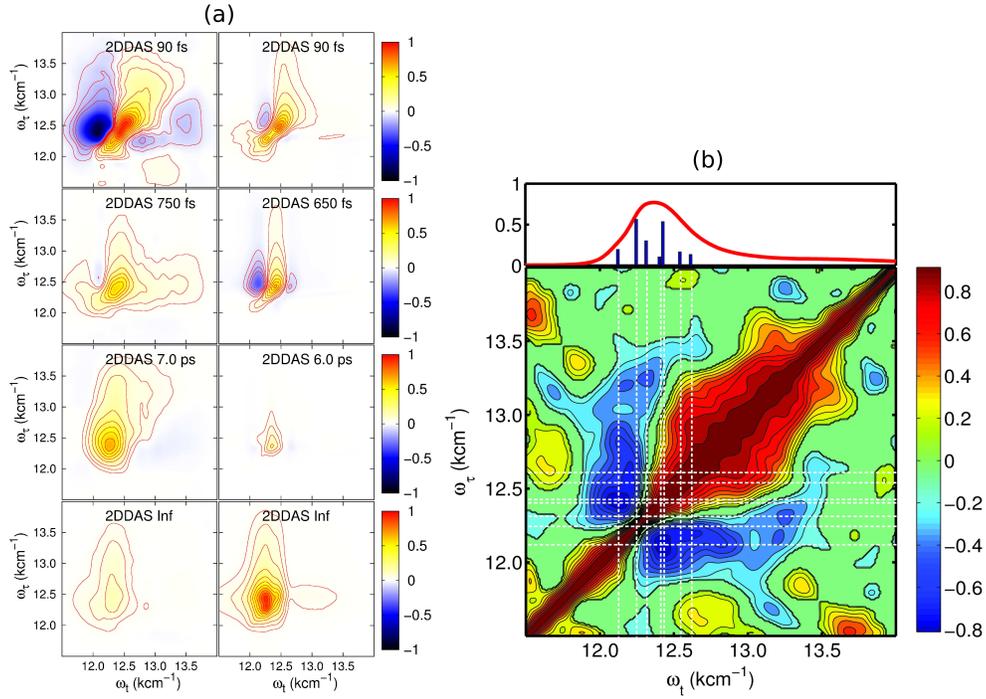

Figure 2: **Analysis of the FMO spectra:** (a) 2D decay associated spectra (2DDAS). The left column shows the experimental results and the right column the theoretically calculated spectra. The four resulting associated decay times $\tau_{1,...,4}$ are indicated in the panels. (b) 2D correlation map of residuals obtained from the series of experimental spectra after subtracting the kinetics by the global fitting procedure. The red line on top is the measured absorption spectrum of the FMO trimer, and the blue bars mark the stick spectrum of the FMO model. The white dashed lines mark the exciton energies, which are used to overlap with the correlation map.

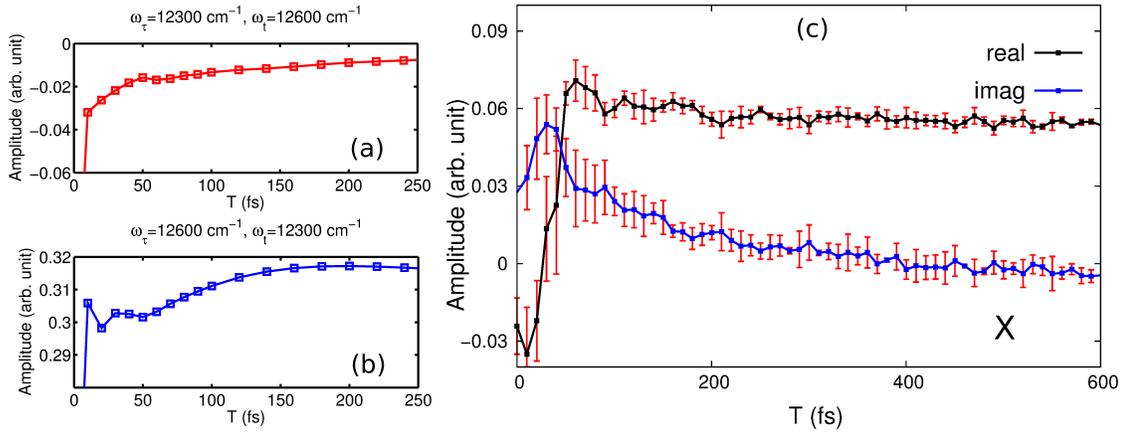

Figure 3: **Time dependence of off-diagonal signals:** Time evolution of the real part of the calculated 2D photon echo signal at the spectral positions (a) $\omega_\tau = 12300$ cm$^{-1}$, $\omega_t = 12600$ cm$^{-1}$ marked by a red square in Fig. 1 (a), and (b) for $\omega_\tau = 12600$ cm$^{-1}$, $\omega_t = 12300$ cm$^{-1}$ marked by a blue square in Fig. 1 (a). It is apparent that a minimal amount of electronic coherence only survives up to $\sim 60$ fs. (c) The real (black), and imaginary (blue) part of the experimentally measured time trace at the same spectral position (see black cross in Fig. 1 (a)) $\omega_\tau = 12350$ cm$^{-1}$, $\omega_t = 12200$ cm$^{-1}$ as measured in Ref. 2, however, measured here at 296 K. The error bars indicate the standard deviation obtained after averaging four data sets. The imaginary part is vertically shifted by $0.035$ for clarity.